\documentclass[prb,amsmath,amssymb,showpacs,showkeys,twocolumn]{revtex4-2}
\usepackage{graphicx}
\usepackage{dcolumn}
\usepackage{rotating}
\usepackage{amsthm,amsmath,amssymb,amsfonts,bbm}
\usepackage{mathptmx}
\usepackage{mathrsfs}
\usepackage{xcolor}
\usepackage{siunitx}
\usepackage{bm}
\usepackage{ulem}
\begin{document}

\title {On dynamical ordering in a 2D electron crystal confined in a narrow channel geometry}

\author{Shan Zou}
\email[E-mail:]{shan.zou@oist.jp}
\affiliation{Quantum Dynamics Unit, Okinawa Institute of Science and Technology (OIST) Graduate University, Onna, 904-0495 Okinawa, Japan}
\author{David G. Rees}
\affiliation{EeroQ Corporation, Lansing, Michigan, 48906, USA}
\author{Denis Konstantinov}
\affiliation{Quantum Dynamics Unit, Okinawa Institute of Science and Technology (OIST) Graduate University, Onna, 904-0495 Okinawa, Japan}

\begin{abstract}

We present both time-averaged and time-resolved transport measurements of a two-dimensional electron (Wigner) crystal on the surface of superfluid helium confined in a narrow microchannel. We find that the field-current characteristics of the driven crystal obtained by the time-averaged measurements exhibit oscillations and negative differential conductivity. This unusual transport behavior was observed previously by Glasson {\it et al.} [Phys. Rev. Lett. \textbf{87}, 176802 (2001)] and was attributed to a nonequilibrium transition of the electron system to a novel dynamically ordered phase of current filaments aligned along the channels. Contrarily to this explanation, our time-resolved transport measurements reveal that oscillating field-current characteristics appear due to dynamical decoupling (slipping) and recoupling (sticking) of the uniform electron crystal to the liquid helium substrate. Our result demonstrates that this unusual non-linear transport effect is intrinsic, does not depend on the device geometry, and is associated with the dynamical interaction of Wigner crystal with the surface excitations of the liquid substrate.              

\end{abstract}

\date{\today}


\maketitle

\section{Introduction}

There is significant interest in understanding physics of systems driven out of equilibrium, and in particular phase transitions between specific types of nonequilibrium states and their universality classes~\cite{Odor2004}. Such questions can be addressed experimentally by finding a suitable model system, such as a collection of interacting particles driven over random or periodic substrates where the system parameters, such as the strengths of interaction and substrate coupling, can be controlled. In recent years many such systems have been experimentally realized, for example superconducting vortex systems~\cite{Blat1994}, charge-density waves~\cite{Grun1988}, colloidal systems~\cite{Murr1987,Murr1989,Murr1990,Marc1996}, and magnetically induced Wigner crystals (WC) in semiconductors~\cite{Andr1988,Zhu2010,Liu2014,Zhan2014}. In such systems, for a weak driving force, the externally driven particles can be pinned by the substrate disorder or by a periodic potential imposed on the system, while the particles can slide off the pinning centers for a sufficiently strong driving. The nonequlibrium sliding state can exhibit a rich variety of dynamic phases depending on the strength of driving and coupling to the pinning centers, which may or may not preserve the symmetry of the equilibrium system~\cite{Reich2017}. 

Electrons trapped on the surface of liquid helium present the cleanest two-dimensional (2D) Coulomb system that can be found in nature~\cite{Konobook}. The ground state of the system is a classical 2D Wigner crystal, which is realized at moderately low temperatures around 1 K. Unlike solid-state charge systems, such as 2D electron gases in semiconductors, electrons on bulk liquid helium are free from disorder. On the other hand, the Wigner crystal can be strongly coupled to a commensurate deformation of the liquid helium substrate, the dimple lattice (DL), which is formed by the pressure exerted on the surface from the spatially ordered electrons localized at the lattice sites. Coupling between 2D Wigner crystal and the DL can lead to significant enhancement of the electron effective mass, thus altering the transport properties of electrons as they are driven along the surface by an external electric field. One of the most remarkable transport phenomena in this system is the saturation of the electron velocity with increasing driving field, which is observed as a plateau in the measured current-voltage characteristics. At the plateau, the velocity of the driven Wigner crystal is limited by resonant emission of the surface capillary waves, ripplons, whose wave vector coincides with the reciprocal-lattice vector of the Wigner crystal. Such ripplons interfere constructively, an effect known as Bragg-Cherenkov (BC) ripplon scattering, thus enhancing the dimple depth and the electron effective mass~\cite{Dykm1997,Mona2009}. The electron system is then dynamically pinned to the surface deformation of the liquid substrate; however it can decouple from DL at sufficiently strong driving fields. This results in an abrupt increase in the measured electron current~\cite{Shir1995}.     

Original experiments on the BC ripplon scattering and the depinning of the Wigner crystal from the surface deformation were performed in a circular Corbino geometry in the presence of a magnetic field applied perpendicular to the surface~\cite{Kris1996,Shir1995}. Later, it was demonstrated that confining electrons in capillary-condensed microchannel devices provides many advantages for experimental studies of such charged systems~\cite{Glas2001,Brad2011,Rees2011,Ikeg2012,Badr2019}. In such devices, the transport of the Wigner crystal at zero magnetic field can be studied by driving it through a long ($\sim 100~\mu$m) and narrow ($\sim 10~\mu$m) microchannel where the electron density and driving electric field are kept uniform along the channel. The field-velocity dependence of the driven electron system can be extracted from the measured current-voltage characteristic of the device, which is typically obtained by averaging over many driving cycles using a conventional lock-in amplifier. An intriguing result was reported by Glasson {\it et al.} who studied the transport of the Wigner crystal confined in a long 16-$\mu$m wide channel at temperatures below 1 K~\cite{Glas2001}. The extracted field-velocity characteristics exhibited an oscillating behaviour and regions of negative differential conductivity, which the authors interpreted as signatures of a nonequilibrium phase transition between the crystalline phase and an anisotropically ordered phase of the moving electron system. The latter phase was envisioned as consisting of spatially ordered current filaments along the edge of the channel, reminiscent of the smectic phase observed in liquid crystals and other systems~\cite{Reich2017}. The oscillating behavior was interpreted as arising from the consecutive formation of current filaments at the edge, and the corresponding depletion of the crystalline phase away from the edge, with increasing driving field. However, the existence of such a novel dynamically ordered phase was questioned by Ikegami {\it et al.} who found the appearance of the effect to be dependent on the sample geometry, and thus concluded that the oscillating behavior of the field-velocity characteristic is not intrinsic but artificial~\cite{Ikeg2009}.  

Here, we present a new experimental study of this unusual effect by employing both time-averaged and time-resolved transport measurements of a 2D Wigner crystal confined in a 20-$\mu$m wide microchannel. First, we confirm that field-velocity characteristics similar to those reported by Glasson {\it et al.} also appear for the same sample geometry as used by Ikegami {\it et al}. This confirms that the effect reported by Glasson {\it et al.} is an intrinsic one and does not depend on the sample geometry. Second, in order to elucidate the intrinsic mechanism of the oscillating field-velocity characteristics, we perform time-resolved measurements over a single driving cycle. The real-time traces of the electron current reveal clear signatures of the repetitive decoupling (slipping) and recoupling (sticking) of the driven Wigner crystal from/to the surface deformation during a single driving cycle. The correspondence between the time-averaged and time-resolved measurement results unequivocally demonstrates that the oscillating field-voltage characteristics arise due to the repetitive stick-slip motion of the driven Wigner crystal, a consequence of its interaction with the liquid substrate. Thus, we conclude that so far there is no experimental evidence for the existence of any novel dynamic phases in this system.                                    
 
\section{Experiment}

\begin{figure}[htt]
\includegraphics[width=0.48\textwidth]{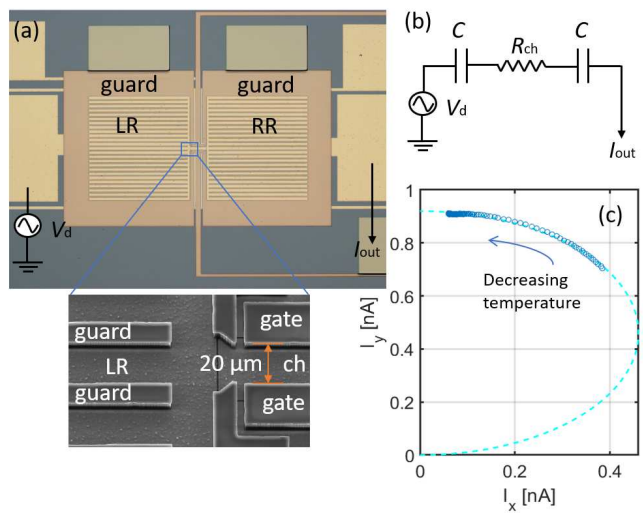}
\caption{(color online) (a) Optical microscope image of the microchannel device. The driving ac voltage $V_\textrm{d}$ is applied to the bottom electrode of the left reservoir (LL) and is thereby coupled capacitively to the electrons on helium, while the current $I_\textrm{out}$ induced by the moving electrons is picked up at the bottom electrode of the right reservoir (RR). The inset shows a scanning electron micrograph of a segment of the 20-$\mu$m central channel adjacent to the left reservoir. (b) Simplified lumped-circuit model of the device. Here,  $C$ is the coupling capacitance between each of the reservoir electrodes and electrons, while the device resistance $R_\textrm{ch}$ mostly comes from the resistance of electrons in the central channel. (c) In-phase ($I_x$) and quadrature ($I_y$) components of the measured current $I_\textrm{out}$ measured for different values of temperature in the range from 0.7 to 1.12~K. The electron resistivity decreases with decreasing temperature. The dashed line represents the fitting to the lumped-circuit model with $C=0.6$~pF.}
\label{fig:1}
\end{figure}

Fig.~\ref{fig:1}(a) shows optical and scanning electron microscope images of the microchannel device used in our experiment. The device consists of two identical arrays of $\SI{20}{\mu m}$-wide and $\SI{700}{\mu m}$-long microchanels, which serve as electron reservoirs, connected by a single $\SI{20}{\mu m}$-wide and $\SI{100}{\mu m}$-long central channel. This structure was fabricated on a silicon oxide substrate using UV lithography methods and is composed of three layers. The bottom gold layer consists of three biasing electrodes which define the bottoms of the left reservoir (LL), right reservoir (RR) and center channel. The corresponding electrical potentials applied to these electrodes are denoted as $V_\textrm{LR}, V_\textrm{RR}$ and $V_\textrm{ch}$, respectively. The top gold layer consists of two confining electrodes, the split-gate and guard electrodes, with corresponding applied potentials denoted as $V_\textrm{ga}$ and $V_\textrm{gu}$. The top and bottom layers are separated by hard-baked photoresist (Allresist ARN 4550) of thickness $h=\SI{4}{\mu m}$. The reservoir microchannels and the central channel are defined by the structure of the guard and split-gate electrodes, respectively, and the photoresist was removed from within the microchannels to form the rectangular-shaped groves of height $h$. The device was placed in a leak tight cell and cooled down at the mixing chamber of a dilution refrigerator. Liquid helium-4 was condensed into the cell and filled the groves by capillary action. 

The surface of the liquid in the microchanels was charged with electrons produced by thermal emission from a tungsten filament placed about 2~mm above the device while positive biases of 0.3~V were applied to reservoir electrodes. The transport of electrons through the central channel was measured by the capacitive-coupling (Sommer-Tanner) method~\cite{Somm1971}, the lumped-element circuit of which is shown in Fig.~\ref{fig:1}(b). Here, $C$ indicates the coupling capacitance between the charged surface and each of the two reservoir electrodes, while $R_\textrm{ch}$ corresponds to the charged surface resistance. Because the area of the reservoirs is significantly larger than that of the central channel, this resistance mostly comes from the electrons in the central channel. An ac ($\omega/2\pi =20~\mbox{kHz}$) voltage $V_\textrm{d}$ was applied to the LR electrode, while the output image current $I_\textrm{out}$ induced by the flow of surface electrons between the reservoirs was measured at the RR electrode by a standard lock-in amplifier which, as in the experiments by Glasson {\it et al.} and Ikegami {\it et al.}, detects the first harmonic component of the ac current response. As an example, the in-phase ($I_x$) and quadrature ($I_y$) components of the current measured at different temperatures in the range from 0.17 to 1.12 K are plotted in Fig.~\ref{fig:1}(c) by opened circles. As the resistance of the electrons in the central channel decreases with decreasing temperature, the resistive ($I_x$) component of the current changes and the current response approaches a purely capacitive-coupling limit ($R_\textrm{ch}\rightarrow 0$). The dashed line shows fitting of the measured current response to the lumped-cicuit model of Fig.~\ref{fig:1}(b), which provides us with an estimate for the coupling capacitance $C=0.6$~F and the resistance $R_\textrm{ch}$ of the electron system in the central channel. The latter varies between 0.7 and 5.9~M$\Omega$ for the experimental data shown in Fig.~\ref{fig:1}(c).   

The conventional transport measurements using a lock-in amplifier have a significant drawback. Because the lock-in amplifier measures the root-mean-square (rms) amplitude of the first harmonic component of the input signal at a given frequency of the ac driving, the lock-in output only coincides with the true rms value when the current response is purely sinusoidal. Alternatively, time-resolved measurements can provide us with more details about complex response of our device. In our experiment, the time-resolved current response over several cycles of the ac driving was measured using a room-temperature fast current preamplifier (Femto DHPACA-100) connected to the RR electrode by a semi-rigid Nb/stainless-steel coaxial cable with the total capacitance of about $300~\mbox{pF}$. This capacitance formed a low-pass filter with the input impedance of the preamplifier, which limited the bandwidth of the measurement circuit to about 1~MHz. The time-resolved current traces were recorded by a digital storage oscilloscope (LeCroy 625Zi) and averaged over several thousand cycles to improve the signal-to-noise ratio. A small current component due to the cross talk between the bottom electrodes was separately measured and subtracted from the averaged traces. 
 
\section{Results}

\begin{figure}[htt]
\includegraphics[width=0.5\textwidth]{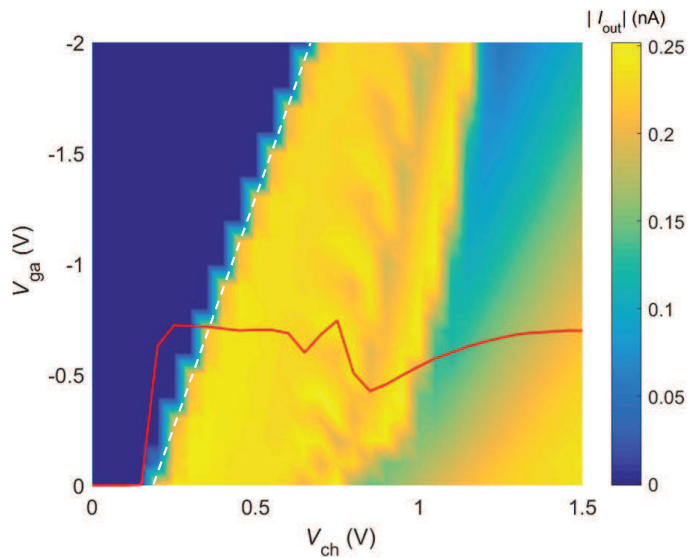}
\caption{(color online) Color map of the current magnitude $|I_\textrm{out}|$ versus $V_\textrm{ch}$ and $V_\textrm{ga}$ obtained at $T=150$~mK using the driving ac voltage $V_\textrm{d}=5$~mV. The channel and gate voltages were scanned with the steps of $V_\textrm{ga}=0.2 \mbox{V}$ and $V_\textrm{ch}=0.02 \mbox{V}$, respectively. The red line shows $|I_\textrm{out}|$ versus $V_\textrm{ch}$ at a fixed value of $V_\textrm{ga}=0$. The white dashed line shows the channel opening boundary.}
\label{fig:2}
\end{figure}

In order to characterize the performance of the device, we first measured the dependence of the transport of electrons through the central channel for different values of the electron density $n_s$ in the central channel. The latter can be controlled by the confining potentials $V_\textrm{ch}$ and $V_\textrm{ga}$ applied to the channel and gate electrodes, respectively. Fig.~\ref{fig:2} shows the magnitude of the current $I_\textrm{out}$ measured for different values of $V_\textrm{ch}$ and $V_\textrm{ga}$ at temperature $T=150$~mK and using a driving voltage amplitude $V_\textrm{d}=5$~mV (note that all ac parameters are given in rms units). Since the number of electrons in the reservoir is much larger than in the central channel and the total number of electrons is conserved, the electrical potential $V_e$ at the charged surface of liquid can be considered independent of $V_\textrm{ch}$ and $V_\textrm{ga}$. From a parallel-plate capacitance model, it is straightforward to obtain the following expression for the electron density in the central channel~\cite{Lin2019}   

\begin{equation}
n_s=\frac{\epsilon\epsilon_0}{\alpha e h} \left( \alpha V_\textrm{ch} + \beta V_\textrm{ga} - V_e \right),
\label{eq:1}
\end{equation}          

\noindent where $\epsilon=1.056$ is the dielectric constant of liquid helium, $\epsilon_0=8.85\times 10^{12}$~F/m is the permittivity of free space, and $\alpha$ and $\beta$ are the weighted contributions ($\alpha +\beta =1 $) to the total capacitance between the electrons and the channel and gate electrodes, respectively. Thus, our device acts essentially as a field-effect transistor (FET) where conductance between the two reservoirs is controlled by the density of electrons in the central channel, and therefore by the voltages $V_\textrm{ch}$ and $V_\textrm{ga}$. The conductance threshold is determined by the condition $\alpha V_\textrm{ch}+\beta V_\textrm{ga}=V_e$ (no electrons in the central channel) which is indicated by the white dashed line in Fig.~\ref{fig:2}. This provides us with an estimate $\alpha/\beta=5$ for our device.        
 
For an opened channel ($\alpha V_\textrm{ch}+\beta V_\textrm{ga}>V_e$) the electron density $n_s$, therefore the conductance between the reservoirs, varies with both $V_\textrm{ch}$ and $V_\textrm{ga}$. As an example, the red solid line in Fig.~\ref{fig:2} shows the variation of current measured for different values of $V_\textrm{ch}$ at a fixed value of $V_\textrm{ga}=0$. Above the conductance threshold, the current magnitude rapidly increases with increasing $V_\textrm{ch}$ as electrons start filling the central channel. For higher values of $V_\textrm{ch}\gtrsim 0.6$~V, the current magnitude decreases due to formation of the Wigner crystal and its dynamical pinning to resonantly emitted ripplons, as described above. However, in contrast to the conductance threshold, it is difficult to determine the phase boundary between electron liquid and Wigner crystal precisely from such measurements. As will be shown below, the decoupling of the Wigner crystal from the ripplons can occur at different times during the driving cycle, depending on a number of parameters including bias voltages and the driving voltage amplitude. This gives rise to the complicated current response to variations in $V_\textrm{ch}$ and $V_\textrm{ga}$ close to the phase boundary in Fig.~\ref{fig:2}.

\begin{figure}[htt]
\includegraphics[width=0.5\textwidth]{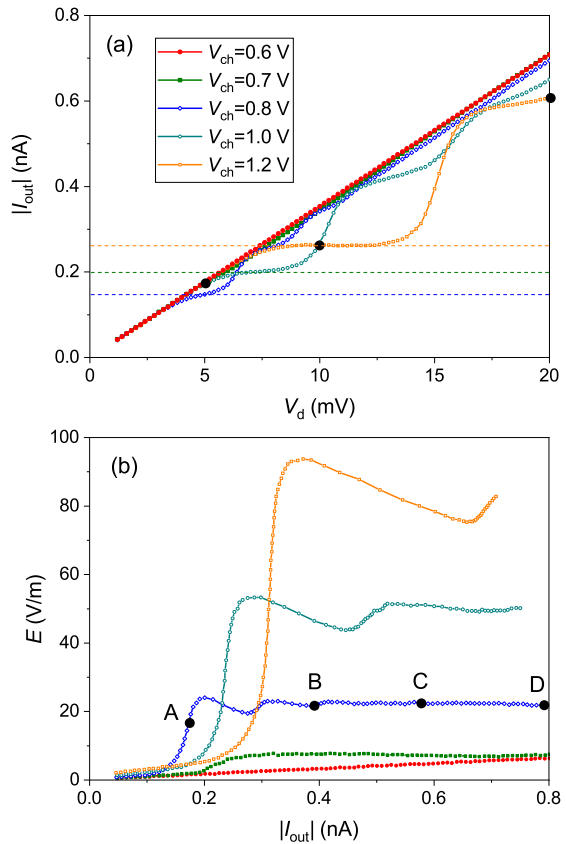}
\caption{(color online) Results of time-averaged transport measurements. (a) Current magnitude $|I_\textrm{out}|$ versus driving voltage $V_\textrm{d}$ obtained at $T=150$~mK, and several values of the channel voltage $V_\textrm{ch}=0.6$ (solid red circles), 0.7 (solid green squares), 0.8 (open blues diamonds), 1.0 (open dark cyan circles), and 1.2 V (open orange squares). The solid black circles on the $IV$-curve for $V_\textrm{ch}=1.2$~V indicate three values of $V_\textrm{d}$ for which real-time traces are presented in Fig.~\ref{fig:4}(a). The dashed lines indicate the BC plateaus corresponding to electron densities $n_e=5.4$, $8.2$ and $\SI{13e8}{cm^{-2}}$, as described in the text.
(b) Corresponding plots of the field-current characteristics for each value of the channel voltage used in panel (a). The in-plane driving electric field $E$ acting on the electrons is extracted from the data shown in panel (a) as described in the text. Solid black circles marked as A, B, C and D on the field-current curve for $V_\textrm{ch}=0.8$~V indicate four values of $V_\textrm{d}=5$, 10, 15 and 20~mV, respectively,  for which real-time traces in Fig.~\ref{fig:4}(b) are presented.}
\label{fig:3}
\end{figure}

\begin{figure}[htt]
\includegraphics[width=0.48\textwidth]{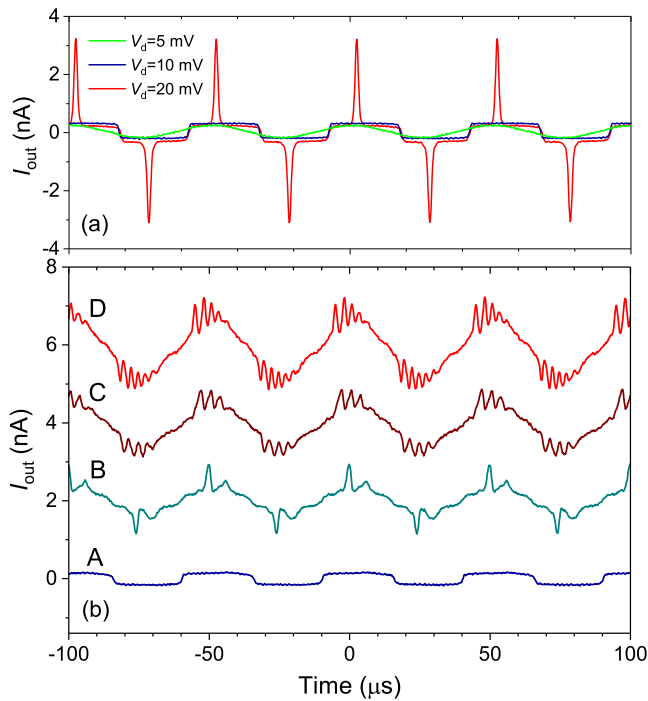}
\caption{(color online) Results of time-resolved measurements. Current $I_\textrm{out}$ versus time recorded over 4 cycles of the ac driving for the channel voltage (a) $V_\textrm{ch}=1.2$~V and (b) $V_\textrm{ch}=0.8$~V and several different values of the driving amplitude $V_\textrm{d}$. The traces in (b) marked as A, B, C and D correspond to $V_\textrm{d}=5$, 10, 15 and 20~mV, respectively. All other conditions are the same as for the data shown in Fig.~\ref{fig:3}.}
\label{fig:4}
\end{figure}

Important information about the dynamical behaviour of surface electrons in different phases can be obtained by measuring $I_\textrm{out}$ as a function of the driving voltage amplitude $V_\textrm{d}$ (I-V curve). The red solid circles in Fig.~\ref{fig:3}(a) show such an I-V curve measured at a fixed value of $V_\textrm{ch}=0.6$~V (note that the gate voltage $V_\textrm{ga}$ is fixed at zero for all data shown in Figs.~\ref{fig:3} and \ref{fig:4}). The current varies linearly with $V_\textrm{d}$, which indicates that the central-channel electrons are presumably in the liquid phase. Contrarily, for higher channel voltages the current is nonlinear and exhibits a series of flat plateaus and abrupt rises. Such a step-like structure becomes more pronounced with increasing $V_\textrm{ch}$ and exhibits longer plateaus with varying $V_\textrm{d}$. In particular, the I-V curve at $V_\textrm{ch}=1.2$~V becomes reminiscent of the usual I-V curve obtained for surface electrons in the solid phase. As described above, it is well established that the strongly nonlinear transport of Wigner crystal on liquid helium arises due to its dynamical interaction with a commensurate lattice of surface dimples comprised of resonantly emitted ripplons. Conventionally, we observe three regimes of such transport~\cite{Kris1996,Shir1995,Ikeg2012,Lin2019}. At sufficiently low driving voltages, the dimples are small and the mobility of the Wigner crystal is close to that of the electron liquid. As the driving voltage increases, the driven system enters the Bragg-Cherenkov scattering regime. The dimples grow and enhance the effective mass of the electrons as their velocity approaches the phase velocity $v_\textrm{ph}$ of the resonant ripplons. Correspondingly, the current saturates at the plateau value of $I_\textrm{BC}=eWv_\textrm{ph}n_s=eW(\sigma/\rho)^{1/2}(8\pi^2 /\sqrt{3})^{1/4}n_s^{5/4}$, where $\sigma$ and $\rho$ are the surface tension and density of liquid helium, respectively, $e$ is the elementary charge, and $W$ is the effective width of the electron system in the central channel. Finally, at a sufficiently high driving voltage the electron crystal decouples (slides) from the dimple lattice. According to a simple phenomenological argument, the sliding of the electron system from the dimples happens when the driving force exceeds the restoring force on an electron from the dimple. In practice, it is very difficult to define the microscopic mechanism of sliding and the corresponding sliding threshold, which depends on many experimental parameters~\cite{Shir1995}.

In general, the I-V curves measured in our device and shown in Fig.~\ref{fig:3}(a) exhibit a more complicated structure than the conventional one described above. The observed multi-step structure, in contrast to a single BC plateau, might suggest a different dynamical state of the electron system. It is convenient to represent these data in the form of the field-velocity characteristics. Following Ref.~\cite{Glas2001}, we can express the time-averaged driving electric field acting on electrons in the central channel as $E=\cos\phi V_\textrm{d}/L$, where $L=\SI{100}{\mu m}$ is the length of the channel and $\phi$ is the phase between $I_\textrm{out}$ and $V_\textrm{d}$. The average electron velocity can be considered proportional to the current magnitude according to $v=|I_\textrm{out}|/(n_seW)$. The field-current curves extracted from data presented in Fig.~\ref{fig:3}(a) are shown in Fig.~\ref{fig:3}(b). It is clear that the multi-step structure of the I-V curves shown in Fig.~\ref{fig:3}(a) results in the oscillating behaviour of the corresponding field-current characteristics. Remarkably, our result is very similar to the field-velocity dependence reported by Glasson {\it et al.} who interpreted it as evidence for a nonequilibrium phase transition between the Wigner crystal and a novel anisotropically ordered phase~\cite{Glas2001}. By this interpretation, the oscillations and flattening of the field-current dependence correspond to the formation of electron current filaments oriented along the edge of the electron system and the corresponding depletion of the crystalline phase at the center, reminiscent of the phase transition in a gas-liquid mixture maintained at constant pressure.     
  
It is important to mention the difference between the geometries of our device and that used in Ref.~\cite{Glas2001}. In our device, the microchannel arrays which comprise the two reservoirs are aligned parallel to the central channel, see Fig.~\ref{fig:1}(a). In the device used by Glasson {\it et al.} the microchannels of the two reservoirs, which were also connected in parallel with each other, were aligned perpendicular to the central channel~\cite{Glas2001}. Both geometries were tested in the experiments reported by Ikegami {\it et al.}, who observed step-like I-V curves only for the perpendicularly-aligned channel geometry similar to that of Ref.~\cite{Glas2001}, while no such behaviour was observed for the parallel channel geometry similar to ours~\cite{Ikeg2009}. Thus, the authors concluded that, since the observed effect depended on the device geometry, it was not associated with electron motion in the central channel but arose due the transport behaviour of electrons in the reservoirs. Nevertheless, in our device we observe exactly the same behaviour of field-velocity dependence as reported in Ref.~\cite{Glas2001}, which demonstrates that the effect is geometry-independent, and therefore intrinsic.

Since the sinusoidal driving electric field acting on the surface electrons alternates with the frequency of tens of kHz, while the intrinsic frequency of the electron-dimple system is typically on the order of tens to hundreds of MHz (the frequency of resonant ripplons), the exact details of the system dynamics are hidden in the current response measured with the lock-in amplifier. Therefore, it is desirable to investigate the time-resolved response of the system during one cycle. Recently, the first time-resolved transport measurements of surface electrons in a microchannel were reported and revealed an interesting stick-slip motion of electrons driven by a linear voltage ramp~\cite{Rees2016}. Motivated by this finding, we performed similar time-resolved transport measurements for electrons in our device under an ac sinusoidal drive. Exemplary results of such time-resolved measurements of $I_\textrm{out}$, taken at $T=150$~mK and the channel voltage $V_\textrm{ch}=1.2$~V (for which conditions the electron system is in the solid state), are shown in Fig.~\ref{fig:4}(a) for three different driving voltage amplitudes $V_\textrm{d}=5$, 10 and 20~mV (see black solid cycles in Fig.~\ref{fig:3}(a)).  It is easy to understand the correspondence between these time-resolved current traces and the current magnitude recorded with the lock-in amplifier plotted in Fig.~\ref{fig:3}(a). For small driving voltage $V_\textrm{d}=5$~mV, the Wigner crystal is in the linear transport regime. Correspondingly, the output current is sinusoidal (green line in Fig.~\ref{fig:4}(a)). For higher driving voltage $V_\textrm{d}=10$~mV, the transport of the Wigner crystal is in the BC scattering regime corresponding to a plateau in the time-averaged I-V curve, see Fig.~\ref{fig:3}(a). Correspondingly, the current saturates at a maximum value given by $I_\textrm{BC}$ during each driving cycle, which results in a nearly rectangular shape of the time-resolved current response (blue line in Fig.~\ref{fig:4}(a)). Finally at sufficiently high driving voltage $V_\textrm{d}=20$~mV, the Wigner crystal slides from the dimple lattice, which results in a significant increase of the current magnitude shown see Fig.~\ref{fig:3}(a). In the corresponding time-resolved current trace (red line in Fig.~\ref{fig:4}(a)) we observe an abrupt rise of current which occurs around the peak value of the ac driving electric field. Soon after this rise the current decreases back to the plateau value $I_\textrm{BC}$, thus producing a narrow spike of current in the time-resolved response. Similar behaviour was observed in Ref.~\cite{Rees2016} for electrons driven by a linear voltage ramp. This has a simple phenomenological explanation. As the Wigner crystal slides from the dimples and the current through the central channel increases, the electrons are transferred rapidly between the two reservoirs, thus decreasing the potential difference between the two opposite ends of the microchannel. As a result, the driving electric field acting on the electrons in the central channel rapidly decreases. When it falls below the sliding threshold, the system re-enters the BC scattering regime, the Wigner crystal is pinned again by the commensurate DL, and the current magnitude returns to the plateau value $I_\textrm{BC}$. In this manner, the Wigner crystal undergoes a stick-slip motion due to dynamical decoupling from, and recoupling to, the commensurate DL~\cite{Rees2016}.                     

Upon establishing the correspondence between the time-averaged and time-resolved measurement results, we now examine the time-resolved current traces corresponding to the multi-step I-V curves shown in Fig.~\ref{fig:3}(a,b). Fig.~\ref{fig:4}(b) show examples of such current traces obtained at $T=150$~mK and the channel voltage $V_\textrm{ch}=0.8$~V for four different driving voltage amplitudes $V_\textrm{d}=5$, 10, 15 and 20~mV (see black solid cycles marked by A, B, C and D in Fig.~\ref{fig:3}(b)). For $V_\textrm{d}=5$~mV (the trace marked by A in Fig.~\ref{fig:4}(b)), the Wigner crystal is in the BC scattering regime, the current is saturated at the plateau value $I_\textrm{BC}$, and the time-resolved current response has a  trapezoidal shape as described above. For $V_\textrm{d}=10$~mV, which corresponds to the second step in the time-averaged curve (marked by B in Fig.~\ref{fig:3}(b)), the time-resolved current trace reveals two stick-slip peaks of current due to a repeated decoupling-recoupling process. For still larger values of $V_\textrm{d}=15$ and 20~mV, the field-current characteristic corresponds to the region of constant field, while the I-V curve is almost linear, see Fig.~\ref{fig:3}(b) and Fig.~\ref{fig:3}(a), respectively. Remarkably, the corresponding time-resolved current traces clearly show repetitive stick-slip motion of the Wigner crystal, which undergoes three and four cycles of the decoupling-recoupling process during one ac driving cycle (traces marked as C and D, respectively, in Fig.~\ref{fig:4}(b)).          

\section{Discussion}

Our time-resolved current measurements provide compelling evidence that the unusual transport behavior of the Wigner crystal reported by Glasson {\it et al.}, and interpreted as evidence for a nonequilibrium phase transition in a driven electron system, in fact arises from the dynamical interaction of the electron lattice with the surface dimples on the liquid substrate. For higher channel voltages, which correspond to both higher electron densities and larger pressing electric fields acting on the electrons in the direction perpendicular to the liquid surface, we expect stronger coupling between the electrons and the dimples. In turn, we expect that the sliding of the Wigner crystal from the DL should occur at higher driving field thresholds~\cite{Shir1995}. The density can be estimated from the value of the plateau current $I_\textrm{BC}$ using the expression given above. As an example, three values of $I_\textrm{BC}$ corresponding to plateaus on the I-V curves for $V_\textrm{ch}=0.8$, 1.0 and 1.2~V are indicated in Fig.~\ref{fig:3}(a) by dashed lines, from which we estimate the corresponding density $n_s=5.4$, 8.2 and $\SI{13e8}{cm^{-2}}$, respectively. Note that these values agree reasonably well with estimates of $n_s$ made using Eq.~\ref{eq:1}. For the largest value of $V_\textrm{ch}=1.2$~V used in our experiment, we observe a plateau and a rise in current similar to the conventional result~\cite{Kris1996,Shir1995,Ikeg2012,Lin2019}. Nevertheless, there is an indication of a second plateau at sufficiently large driving voltages $V_\textrm{d}\gtrsim 15$~mV (see Fig.~\ref{fig:3}(a)), and it is therefore possible that the second sliding event would occur at still higher $V_\textrm{d}$. The repetitive pinning-sliding becomes clearly evident at lower values of $V_\textrm{ch}$ where the coupling between the electrons and the dimples is weaker. We note that multiple decoupling was also observed in Ref.~\cite{Rees2016} by increasing temperature $T$ towards the value $T_m=e^2\sqrt{\pi n_s}/(4\pi\epsilon_0k_B\Gamma)$ ($\Gamma\approx 127$ is the plasma parameter~\cite{Grim1979,Dykm1997a}) corresponding to the melting temperature of the Wigner crystal. This agrees with our result because the coupling strength and decoupling threshold will also decrease with increasing $T$~\cite{Shir1995}. Interestingly, the melting temperature of 150~mK would correspond to an electron density $n_s=\SI{0.5e8}{cm^{-2}}$, which is significantly lower than the estimated density for $V_\textrm{ch}\gtrsim 0.6$~V. This suggests that the electron system is already in the solid phase at $V_\textrm{ch}=0.6$~V, despite its linear response across the range of driving voltages used in the experiment. This could be due to a continuous and rapidly repeated coupling-decoupling process, in the limit of which the Wigner crystal is essentially decoupled from the heavy DL and thus exhibits a high mobility comparable to that of electrons in the liquid phase.

Finally, we discuss the relevant timescale of the observed stick-slip transitions. As described above, the characteristic frequency of the DL formation is similar to that of the resonant ripplons $\sim \sqrt{\sigma/\rho}n_s^{3/4}$, which is between 10 and 100~MHz for the typical electron densities achieved in our experiment. Another relevant process is the redistribution of charge between the two reservoirs, which affects the driving electric field acting on the electrons. The typical rate of this process is given by $(R_\textrm{ch}C)^{-1}$, where $C=0.6$~pF is the capacitance between electrons and reservoir electrodes. For a typical resistance $R_\textrm{ch}=700$~k$\Omega$ of the Wigner crystal in the sliding state we estimate the corresponding frequency to be 2.4~MHz. Thus we conclude that the observed response time in our time-resolved current traces is limited by the bandwidth of our experimental setup (about 1~MHz), which is mainly due to the bandwidth of our current preamplifier as well as the low-pass filter formed by its input impedance and the connecting cable, as discussed earlier. As a future improvement of our method, we can consider employment of a cryogenic fast current preamplifier which would allow the length of the connecting cable to be decreased significantly, thus improving the response time of our detection circuit.                          

\section{Summary}

In summary, we studied the unusual transport phenomenon reported earlier by Glasson {\it et al.} in a Wigner crystal of surface-state electrons on liquid helium confined in a microchannel. In our experiment, we employed both time-averaged and time-resolved measurements of the electron current that flows in response to an ac (sinusoidal) driving voltage. Contrary to Glasson {\it et al.}, who interpreted the oscillatory response in the field-velocity dependence as a signature of a novel dynamically ordered many-electron state~\cite{Glas2001}, we conclude that it arises from the dynamical interaction of the Wigner crystal with surface excitations of the liquid helium. In particular, our time-resolved measurements provide compelling evidence that the non-linear features in the field-velocity curves arise due to the repetitive stick-slip motion of the conventional Wigner crystal interacting with the dimple lattice formed on the liquid substrate. Besides the importance of our work for understanding the dynamics of driven electrons on liquid cryogenic substrates, our observations could be interesting in the general context of friction whose microscopic nature presents an interesting and challenging question~\cite{Vano2007,Wang2008,Vano2013}.      
 
{\bf Acknowledgements} The work was supported by an internal grant from Okinawa Institute of Science and Technology (OIST) Graduate University.



\end{document}